\documentclass{osa-article}
\pdfoutput=1
\journal{osajournal}

\articletype{Tutorial}

\begin{document}

\begin{flushright}
QMUL-PH-21-25\\
\end{flushright}

\title{The double copy: from optics to quantum gravity}

\author{Chris D. White,\authormark{1}}

\address{\authormark{1}Centre for Research in String Theory, School of Physics and Astronomy, Queen Mary University of London, 327 Mile End Road, London E1 4NS, UK}

\email{\authormark{*}christopher.white@qmul.ac.uk} 



\begin{abstract}
Recently, an intriguing relationship (the {\it double copy}) has been discovered between theories like electromagnetism, and gravity. This potentially gives us a new way to think about gravity, and there are also practical applications involving the efficient calculation of gravitational observables, and also how to simulate gravity using optical systems. In this tutorial, we will review what is known about the double copy, and argue that now is the perfect time for researchers in optics and / or condensed matter to become interested in this fascinating correspondence.
\end{abstract}

\section{Introduction}

Our current understanding of the ultimate building blocks of nature is in terms of matter, acted on by forces. To the best of our knowledge, there are only four fundamental forces: electromagnetism, the weak and strong nuclear forces, and gravity. The first three of these are described by the Standard Model of Particle Physics, which is a quantum field theory and thus incorporates the effects of both quantum mechanics and relativity. Gravity, on the other hand, is best described by General Relativity, which has so far stubbornly resisted humankind's attempts to make it quantum. This is unfortunate, given that quantum gravitational effects are believed to be important in extreme regions of the universe where GR breaks down e.g. the centre of black holes, or at the Big Bang itself. However, even classical GR is at the cutting edge of fundamental physics research: the recent discovery of gravitational waves by the LIGO experiment~\cite{Abbott:2016blz} provides an entirely new tool with which to view our universe. Typical signals arise from the the inspiral and merger of heavy objects such as black holes and neutron stars, and the precision with which such signals can be understood is limited by our ability to approximate these physical situations in GR calculations. 

The above discussion suggests that new ways of thinking about gravity are useful for two main reasons: (i) they could generate new conceptual insights into what a complete (quantum) theory of gravity might look like; (ii) they might provide clever ways of streamlining difficult calculations, even in the incomplete theory of gravity (GR) that we know about. The aim of this article is to review a particularly intriguing way to think about gravity, namely that it can be seen -- in a well-defined sense that we will make precise -- as two copies of a {\it gauge theory} (the type of theory describing the other forces in nature). This is known as the {\it double copy}, and was first discovered in the so-called {\it scattering amplitudes} that are related to probabilities for particles to interact in collider experiments~\cite{Bern:2008qj,Bern:2010ue,Bern:2010yg}. It has since been extended to other types of solution, and there are even cases where certain quantities in GR can be obtained by "copying" counterparts in conventional electromagnetism. Whilst analogies between electromagnetic and gravitational physics have been made many times before, the double copy constitutes a very precise relationship, that opens up the possibility to simulate and explore gravity using optical or condensed matter systems, in a very systematic way. Furthermore, the double copy between gauge and gravity theories turns out to be only one example of a whole host of such copy relationships, that have been shown to relate quantities in an increasingly complex web of both physical, and non-physical theories (see e.g. ref.~\cite{Bern:2019prr} for a wide-ranging review). This constitutes mounting evidence that our traditional way of thinking about field theories obscures a profound underlying structure, such that the rules of quantum field theory that we take for granted may have to be rewritten. We also do not know if all field theories can be obtained by "copying" other ones, or whether there is a special set of "copiable" theories, that is picked out for some sort of special reason. 

As the scope and range of applications of the double copy have widened, so has the network of scientists who are interested in its consequences. The field of interested people is an interdisciplinary mix of high energy physicists (trained in either particle physics or string theory); astrophysicists and cosmologists; and both pure and applied mathematicians. There is even an entire annual conference devoted to the topic ({\it QCD Meets Gravity}), whose most recent incarnation had over 250 participants from 23 countries! As is common in academia, most of the barriers for knowledge transfer arise from the lack of a common language for what is very often the same physics. Indeed, finding such a language may itself shed light on the origin of the double copy itself. Furthermore, there is clearly scope for researchers from other branches of physics -- such as optics and condensed matter -- to become involved, and to suggest theoretical methods, or experimental set-ups, that can be used to further our knowledge of the double copy. Given that this is a young subject involving unfamiliar territory, there is a high probability of generating novel and impactful research insights by bringing in well-established ideas and techniques from other areas. To define the double copy more precisely, we must first review the various theories that it relates.

\section{Gauge Theories}
\label{sec:gauge}

All of the forces and matter particles in the Standard Model are described by fields, whose equations of motion have wave-like solutions. In the quantum theory, these waves arrive in discrete quanta or lumps, giving rise to the observed matter and force-carrying particles. For example, the force particles associated with the electromagnetic and strong nuclear forces are the {\it photon} and {\it gluon} respectively. The former is described by the quantum version of Maxwell's equations, which in many applications are written in a non-relativistic form involving separate electric fields $\vec{E}$ and $\vec{B}$. In high-energy physics applications, however, it is convenient to use a different language that makes the compatibility of electromagnetism with the theory of Special Relativity manifest (for a textbook review of the following, see e.g. ref.~\cite{jackson_classical_1999}). In SR, space and time mix with each other under the {\it Lorentz transformations} that take us between different inertial frames. It thus makes no sense to separate space and time coordinates, and instead we may combine them to form so-called {\it 4-vectors}: 
\begin{equation}
    x^\mu=(t,\vec{x}),
    \label{4position}
\end{equation}
where we use {\it natural units} such that the speed of light $c=1$ (also Planck's constant $\hbar=1$), and the index $\mu\in\{0,1,2,3\}$ labels which component of the 4-vector we are talking about. The 4-vector in eq.~(\ref{4position}) specifies the position of an event in spacetime, and the vector itself transforms straightforwardly under Lorentz transformations, regarded as $4\times 4$ matrices. Given eq.~(\ref{4position}), it is also convenient to define a vector with the index downstairs:
\begin{equation}
    x_\mu=(t,-\vec{x})=\eta_{\mu\nu}x^\nu.
    \label{xlower}
\end{equation}
That is, the downstairs 4-vector consists of flipping the spatial components relative to the upstairs one, and the right-hand side shows that this operation can be formalised by multiplying the components of the original 4-vector with the {\it metric tensor}, expressed in matrix form as
\begin{equation}
    \eta_{\mu\nu}=\left(\begin{array}{rrrr}
         1 & 0 & 0 & 0 \\
         0 & -1 & 0 & 0\\
         0 & 0 & -1 & 0\\
         0 & 0 & 0 & -1
    \end{array}\right).
    \label{etadef}
\end{equation}
We have also used the {\it Einstein summation convention} in eq.~(\ref{xlower}), where repeated indices are assumed to be summed over. The convenience of the downstairs index notation is that one may define a dot product of two 4-vectors
\begin{equation}
    x\cdot y=x^\mu y_\mu=\eta_{\mu\nu}x^\mu y^\nu,
    \label{dotproduct}
\end{equation}
which turns out to be invariant under Lorentz transformations. Although we have talked about spacetime positions here, any set of four numbers that
mixes appropriately under Lorentz transformations may be combined to make a 4-vector. A further example is the 4-momentum of a particle:
\begin{equation}
    p^\mu=(E,\vec{p}),
    \label{4momentum}
\end{equation}
where $E$ and $\vec{p}$ are the (relativistic) energy and momentum respectively. By evaluating the dot product of eq.~(\ref{4momentum}) with itself in a general frame, and in the rest frame of the particle, one obtains the relativistic energy-momentum relation
\begin{equation}
    E^2-\vec{p}^2=m^2,
    \label{Emom}
\end{equation}
where $m$ is the particle mass. In electromagnetism, it turns out that one may combine the electrostatic potential $\phi$ and magnetic vector potential $\vec{A}$ into a single 4-vector known as the {\it gauge field}:
\begin{equation}
A_\mu=(\phi,\vec{A}),
\label{Amudef}
\end{equation}
where the electric and magnetic fields are defined in our notation by 
\begin{equation}
\vec{E}=-\nabla \phi-\frac{\partial \vec{A}}{\partial t},\quad
\vec{B}=\nabla\times\vec{A}.
\label{EBdef}
\end{equation}
With these definitions, the usual non-relativistic Maxwell equations can be shown, after some effort, to arise from the relativistic equations~\footnote{Actually, eq.~(\ref{Maxwell}) generates only two of the four usual non-relativistic Maxwell equations. The other two arise from the {\it Bianchi identity} $\partial_{[\alpha}F_{\beta\gamma]}=0$, where square brackets denote antisymmetrisation over all indices. This identity is an automatic consequence of the definition of the field strength in eq.~(\ref{Fdef}). }
\begin{align}
    \partial_\mu F^{\mu\nu}&=j^\nu,
\label{Maxwell}
\end{align}
where $\partial_\mu\equiv\partial / \partial x^\mu$, and we have defined the {\it field strength tensor}
\begin{equation}
    F_{\mu\nu}=\partial_\mu A_\nu-\partial_\nu A_\mu,
    \label{Fdef}
\end{equation}
as well as the current density 4-vector
\begin{equation}
    j^\mu=(\rho,\vec{j})
\label{jmudef}
\end{equation}
where $\rho$ and $\vec{j}$ are the charge and (3-vector) current densities respectively. 

The theory of electromagnetism has a remarkable rich symmetry known as {\it gauge invariance}: if we subject $A_\mu$ to the transformation
\begin{equation}
    A_\mu\rightarrow A'_\mu=A_\mu+\partial_\mu \chi,
    \label{gauge}
\end{equation}
where $\chi$ is a (spacetime-dependent) scalar field, the field strength tensor of eq.~(\ref{Fdef}) does not change, and thus the equations of electromagnetism remain invariant. Gauge invariance turns out to be crucial in being able to formulate a sensible quantum theory of electromagnetism, but also means that the gauge field corresponding to a given physical system is not unique, but can be written in infinitely many ways. One may make things precise by imposing additional constraints on $A_\mu$, known as {\it fixing a gauge}. A common gauge choice is the Lorenz gauge 
\begin{equation}
    \partial_\mu A^\mu=0,
    \label{Lorenz}
\end{equation}
for which the first equation in eq.~(\ref{Maxwell}) reduces to 
\begin{equation}
   \left( \frac{\partial^2}{\partial t^2}-\nabla^2\right)A_\mu=j_\mu.
       \label{wave}
\end{equation}
In the vacuum case ($j_\mu=0$), this reduces to the wave equation, whose solutions include the well-known plane waves
\begin{equation}
    A_\mu=\epsilon_\mu e^{ik\cdot x},\quad k_\mu \epsilon^\mu=0,
    \label{planewaves}
\end{equation}
where $\epsilon_\mu$ is the polarisation vector, and $k_\mu$ can be interpreted as the 4-momentum~\footnote{In natural units, the frequency $\omega$ and energy $E$ are interchangeable, as are the 3-momentum $\vec{p}$ and wave vector $\vec{k}$.}. In quantum field theory, particle states (photons) correspond to quanta of these plane waves. 

\begin{figure}[h]
\centering
    \includegraphics[width=0.4\textwidth]{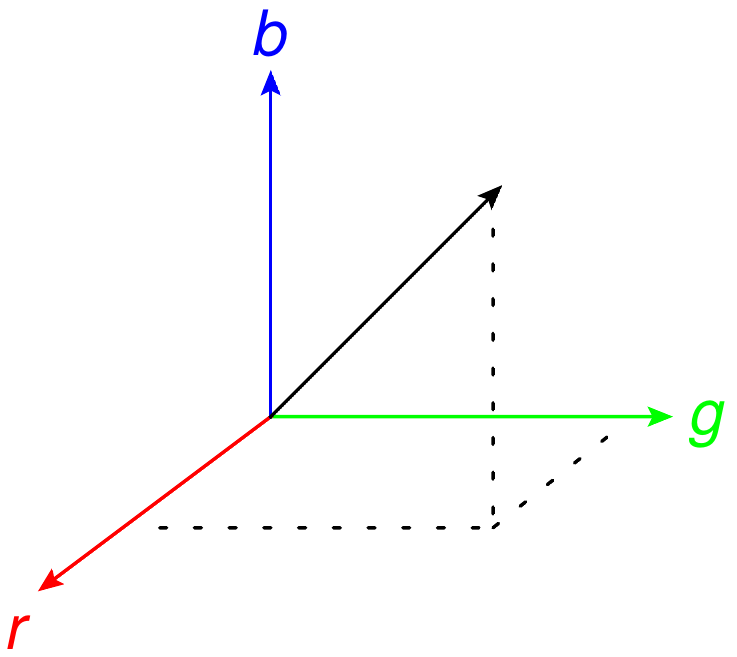}
        \caption{We can imagine the quark field at each point in spacetime as having an arrow in an internal {\it colour space}, telling us how much of each colour charge it has.}
    \label{fig:quarkarrow}
\end{figure}

Above, we saw how gauge transformations act on the photon field $A_\mu$. In quantum electrodynamics, there is also an electron field $\psi(x^\mu)$ which, being complex, has a phase at each point in spacetime~\footnote{The electron field is a type of mathematical quantity called a {\it spinor}, but the details of this are irrelevant to our main discussion.}. A (local) gauge transformation acting on the electron is defined to be a phase shift, which may be defined separately at each point:
\begin{equation}
    \psi\rightarrow e^{i\theta(x^\mu)}\psi.
    \label{abelian_gauge}
\end{equation}
There is a nice geometric way to think about this: the phase at each spacetime point can be represented as an arrow on the unit circle (figure~\ref{fig:quarkarrow}(a)), such that performing a local gauge transformation amounts to rotating this arrow differently at different points in spacetime. 
\begin{figure}[h]
    \centering
    \includegraphics[width=0.4\textwidth]{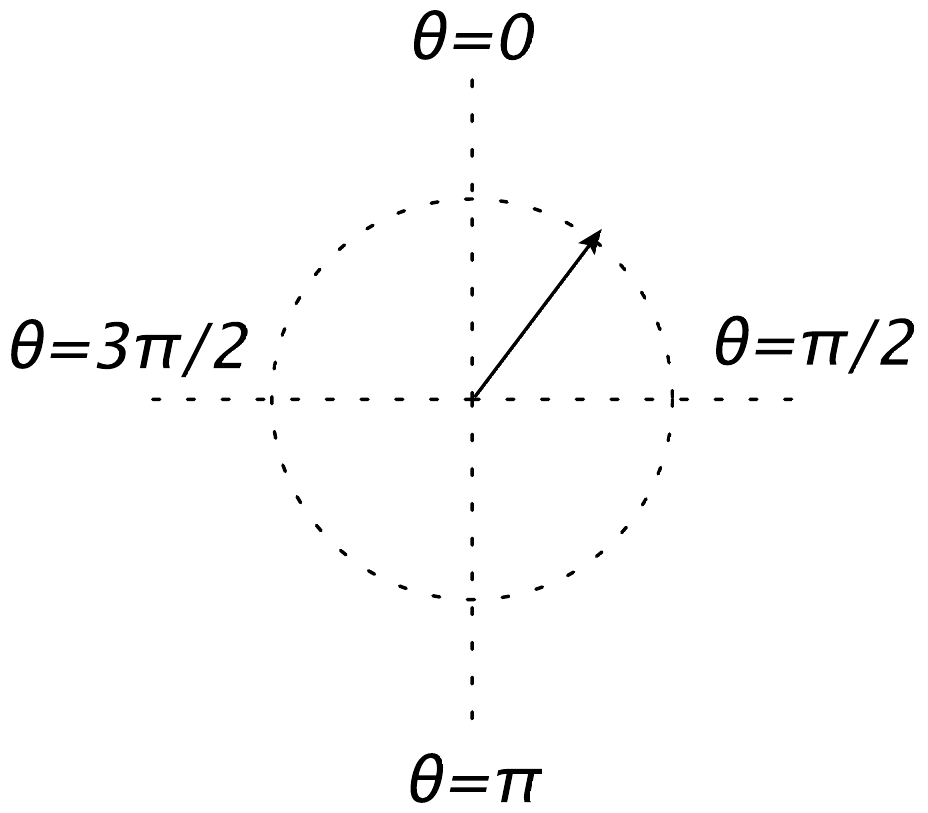}
    \caption{The phase of the electron field at each point in spacetime can be thought of as an arrow on the unit circle, where a local gauge transformation may rotate the arrows at different points by different amounts.}
    \label{fig:phase}
\end{figure}
One may show that requiring that the {\it Dirac equation} for the electron be invariant under these local gauge transformations {\it requires} the presence of the field $A_\mu$ that transforms as in eq.~(\ref{gauge}). Thus, electromagnetism can be completely derived as a consequence of local gauge invariance! The question naturally arises of whether the other forces in nature can be phrased in terms of symmetries. Indeed they can, and we may take as an example the strong force, that acts on fundamental particles called quarks. Quarks carry a type of charge called colour, which can take three different values (conventionally labelled as $r$, $g$ and $b$, or "red", "green" and "blue"). We can then think of the quark field as carrying an abstract "colour arrow" at each point in spacetime, that tells us how much redness, greenness and blueness there is: see figure~\ref{fig:quarkarrow}. 
A local gauge transformation in this case corresponds to rotations of the colour arrow by different amounts at different spacetime points, and demanding that the equation of motion of the quark field be invariant under local rotations of the colour arrow leads to the presence of a gauge field for the gluon, albeit more complicated than the electromagnetic case above. 
The group of all possible colour rotations of the quark field (i.e. rotations in a 3-dimensional complex colour space) is denoted by SU(3), and mathematicians call such sets of continuous transformations {\it Lie groups}. Acting on the quark field, members of the colour rotation group will be a 3$\times$3 complex matrix, and it turns out that any such member can be written as 
\begin{equation}
    {\bf U}=\exp\left[i\theta^a(x^\mu) {\bf T}^a\right],
    \label{Lietheorem}
\end{equation}
where the matrices ${\bf T}^a$ are called {\it generators}, and the parameters $\theta^a(x^\mu)$ label which member of the group we are talking about. Equation~(\ref{Lietheorem}) is known as {\it Lie's theorem}, and applies to any continuous group: note, for example, that the electron phase rotations of eq.~(\ref{abelian_gauge}) have this form, where there is only one parameter $\theta$, and one generator (the number 1, which is not explicit!). For SU(3), it turns out that 8 generators are needed, and thus the index $a$ runs from 1 to 8. The explicit form of the generator matrices depends on the choice of a suitable basis in colour space, but the generators for any Lie group can be shown to obey a so-called {\it Lie algebra}
\begin{equation}
    [{\bf T}^a,{\bf T}^b]=if^{abc}{\bf T}^c,\quad
    [{\bf A},{\bf B}]\equiv {\bf A}{\bf B}-{\bf B}{\bf A}.
    \label{Liealgebra}
\end{equation}
That is, the commutator of any two generators must itself be equal to a superposition of generators~\footnote{The proof of this result comes from demanding consistency between eq.~(\ref{Lietheorem}), and the fact that the product of any two transformations must itself be a transformation in the group.}, where the relevant coefficients $\{f^{abc}\}$ are called {\it structure constants}, are unique for any given Lie group, and have the property of being completely antisymmetric under interchange of any two indices. A familiar example of eq.~(\ref{Liealgebra}) from our undergraduate days is for the group SU(2), which arises in the description of angular momentum. Equation~(\ref{Liealgebra}) is then the statement that a commutator of two angular momentum generators is itself a generator of angular momentum~\footnote{The structure constants for SU(2) turn out to be the Levi-Civita symbol $\epsilon^{abc}$, so that one has e.g. $[J_x,J_y]=iJ_z$ for the angular momentum generators.}. Recall that for the phase shifts of electromagnetism, there was only one generator (the number 1), which clearly commutes with itself! Thus, the structure constants vanish in electromagnetism.

In electromagnetism, requiring invariance under phase rotations -- which had one degree of freedom at each point -- led to a gauge field $A_\mu$. For the strong force, colour rotation invariance of the quark field ends up leading to 8 separate gauge fields, one for each value of the index $a$ above. We may then write this as a single field $A_\mu^a$, where $\mu$ is the spacetime index, and $a$ the colour index. Despite the abstract nature of the present discussion -- which unavoidably arises from the esoteric nature of the symmetries of the quark field, and how these must be expressed mathematically -- it is nevertheless possible to understand the extra complications of gluons with respect to photons in highly physical terms.
It turns out that a quark may change colour by emitting a gluon, such that conservation of colour charge requires that the gluon itself carry colour charge. The index $a$ in the gluon field then simply tells us which type of gluon we are talking about. 

Armed with the above notation, we can finally state the equations describing the gluon. We will focus on the case in which the quark fields are absent, which is known as {\it Yang-Mills theory}. One may then define a field strength tensor
\begin{equation}
F_{\mu\nu}^a=\partial_\mu A_\nu^a-\partial_\nu A_\mu^a
+gf^{abc}A_\mu^b A_\nu^c,
\label{gluonfieldstrength}
\end{equation}
where $g$ is a number called the {\it coupling constant}. It represents the strength of the strong force, and is the analogue of the electron charge $e$ in electromagnetism. In terms of this field strength, the vacuum {\it Yang-Mills equations} are~\footnote{Similar to the electromagnetic case above, we have ignored a Bianchi identity that arises from the definition of the field strength tensor.}
\begin{equation}
\partial^\mu F_{\mu\nu}^a+gf^{abc} A^{\mu b}F^{c}_{\mu\nu}=0.
    \label{YM}
\end{equation}
We see that eqs.~(\ref{gluonfieldstrength}) and~(\ref{YM}) are more complicated than their electromagnetic counterparts of eqs.~(\ref{Fdef}, \ref{Maxwell}). First, they involve the structure constants, which are absent in the electromagnetic case, although their presence here should not surprise us: the symmetry of the theory (as represented by the structure constants) is dictating the equations of motion! Secondly, the Yang-Mills equations are clearly non-linear, in striking contrast to the Maxwell equations. The physical interpretation of this is that these terms represent {\it interactions} between different gluons, which is entirely to be expected given that gluons carry colour charge, as noted above. In electromagnetism, the photon carries no charge, and thus cannot interact directly with itself~\footnote{Photons can mutually interact indirectly, by coupling to an intermediate fermion bubble, but this does not invalidate the above discussion!}. We may further check the consistency of the above equations by showing how electromagnetism emerges in the appropriate limit. Firstly, we see that all of the non-linear terms in eqs.~(\ref{gluonfieldstrength}, \ref{YM}) involve the structure constants $f^{abc}$. The latter vanish for electromagnetism, so that the equations linearise. Secondly, the index $a$ can only take a single value in electromagnetism, so that eqs.~(\ref{gluonfieldstrength}, \ref{YM}) reduce to eqs.~(\ref{Fdef}, \ref{Maxwell}), as required. 

Note that, even in Yang-Mills theory, there are situations where the non-linear terms vanish or can be ignored, such that $A_\mu^a$ (for each $a$) obeys an equation similar to the Maxwell equations. By fixing a gauge, one may obtain the wave equation of eq.~(\ref{wave}). The physical gluon particle then arises as a quantum of a plane-wave solution.

\section{Gravity}
\label{sec:GR}

\begin{figure}
    \centering
    \includegraphics[width=0.3\textwidth]{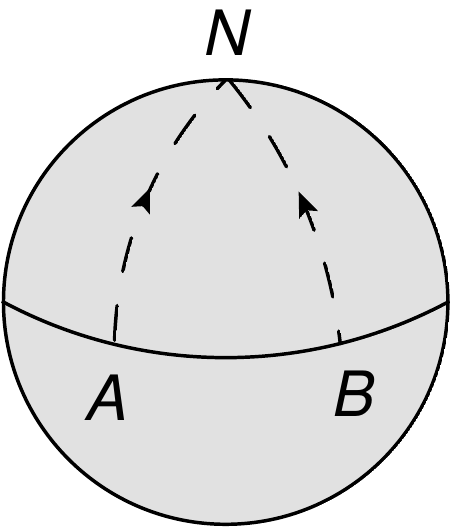}
    \caption{Trajectories of two people walking in a curved space.}
    \label{fig:sphere}
\end{figure}

Having reviewed what a gauge theory is, let us now turn to gravity. This is described by General Relativity, whose basic idea is that matter and energy curve the spacetime that they are sitting in. Freely falling test particles will then follow paths of shortest distance in a curved space, and this distorted motion then corresponds to the force of gravity. As an analogy, consider 
a simple two-dimensional curved space, namely the surface of the sphere in figure~\ref{fig:sphere}. Let us take two people at points $A$ and $B$, and instruct them both to walk towards the north pole $N$. Each person will feel that they are walking locally in a straight line with no forces acting upon them. However, merely the fact that they are walking on a curved surface will lead to them moving closer together, which looks like an attractive force. 

To make the above idea precise, we first need to use the language of 4-vectors to combine space and time. Next, we need to see mathematically how to describe a curved space. First, let us consider spacetime without gravity, as in the previous section. Given the spacetime displacement 
\begin{displaymath}
dx^\mu=(dt,d\vec{x})
\end{displaymath}
between two events, one may take its dot product with itself to form an invariant distance 
\begin{equation}
    ds^2=dt^2-d\vec{x}\cdot d\vec{x}=\eta_{\mu\nu} dx^\mu dx^\nu,
    \label{ds2}
\end{equation}
where we have rewritten things to explicitly involve the metric tensor of eq.~(\ref{etadef}). We thus see that $\eta_{\mu\nu}$ can be thought of as telling us how to measure distances in spacetime, and it is for this reason that it is called the "metric" tensor, which in this particular case is known as the {\it Minkowski metric}. 

If we now add gravity, spacetime will be distorted, where this distortion will potentially be different at different spacetime points. We can easily model this by replacing $\eta_{\mu\nu}\rightarrow g_{\mu\nu}(x^\mu)$ in eq.~(\ref{ds2}), where the dependence of the new metric tensor $g_{\mu\nu}$ on spacetime position indicates that we have warped our previously flat space to make a potentially curved one. Different curved spacetimes will have different metric tensors $g_{\mu\nu}$, and to complete the theory we need equations that tells us how to determine the metric in all of spacetime that corresponds to a given distribution of matter and energy. These are the  {\it Einstein field equations}, and are extremely complicated to solve in general. Famous solutions include the possibility of {\it black holes} (regions of spacetime from which not even light can escape), and universes which expand outwards from a finite time in the past (the Big Bang). 

Note that there are many situations in which we are far away from a given matter distribution, in which case we can try to find an approximate solution for the gravitational field. To do this, one may write the following ansatz:
\begin{equation}
    g_{\mu\nu}=\eta_{\mu\nu}+\kappa h_{\mu\nu},
    \label{hdef}
\end{equation}
where $\kappa=\sqrt{32\pi G_N}$ is a conventional constant factor containing Newton's constant, and $h_{\mu\nu}$ is the {\it graviton field}, which represents the deviation from flat space. Upon substituting eq.~(\ref{hdef}) into the Einstein equations, we can collect terms involving successive powers of the small number $\kappa$, where keeping each additional power amounts to a better approximation to the exact metric tensor describing the system. This is {\it perturbation theory}, and a similar procedure may be used to find classical solutions to the non-linear Yang-Mills theory described in the previous section, where the coupling constant $g$ is the relevant expansion parameter in that case.

In gauge theory, we saw that the field $A_\mu$ is not unique for a given physical system, but varies upon making gauge transformations as in eq.~(\ref{gauge}). Likewise, one may show that the metric tensor of a given spacetime in GR transforms in the following way under coordinate transformations $x^\mu\rightarrow y^\mu$:
\begin{equation}
    g_{\alpha\beta}(y^\mu)=\left(\frac{\partial x^\mu}{\partial y^\alpha}\right)
    \left(\frac{\partial x^\nu}{\partial y^\beta}\right)g_{\mu\nu}(x^\mu).
    \label{gtrans}
\end{equation}
There are infinitely many different coordinate systems we can choose, and thus infinitely many ways of writing the metric tensor for the {\it same} physical system! As in gauge theory, we may fix things by imposing additional constraints on the metric. In perturbation theory, this is usually done at the level of the graviton field itself, and one example is the {\it transverse-traceless (TT) gauge}, in which the graviton satisfies
\begin{equation}
\eta^{\mu\nu}h_{\mu\nu}=0,\quad \partial_\mu h^{\mu\nu}=0.
\label{TT}
\end{equation}
The vacuum Einstein equations then become
\begin{equation}
\left(\frac{\partial^2}{\partial t^2}-\nabla^2\right)h_{\mu\nu}=0,
\label{hwave}
\end{equation}
which can immediately be recognised as the wave equation! The solutions constitute small ripples in the fabric of spacetime, and are precisely the gravitational waves discovered by LIGO. A particularly straightforward set of solutions are the plane waves
\begin{equation}
    h_{\mu\nu}=\epsilon_{\mu\nu}e^{ik\cdot x},
    \label{hplanewave}
\end{equation}
where $\epsilon_{\mu\nu}$ is a polarisation tensor. It turns out that gravitational waves have two polarisation states, as for photons, and the similarity between eqs.~(\ref{planewaves}, \ref{hplanewave}) is our first glimpse of an intimate relationship between gauge theory and gravity~\footnote{The comparison between electromagnetic and gravitational waves has also been explored in ref.~\cite{Barnett:2014era}, in a way that makes the common physics exceptionally clear.}. Quanta of these waves are known as {\it gravitons}, and are the hypothetical particles of quantum gravity.

\section{Scattering amplitudes and the double copy}
\label{sec:amplitudes}

We have seen in the previous section that the wave equation naturally arises in both gauge theories and gravity, where the force-carrying particles in each case arise as quanta of plane wave solutions, generalising the origin of the photon in electromagnetism. In applications of quantum field theory (QFT), a commonly occuring situation is a scattering experiment, in which two (beams of) particles interact, and produce a number of other particles in the final state. In a quantum theory, we cannot predict exactly what will happen, but can instead calculate the probability of given final states. Each scattering process is then associated with a complex number called the {\it scattering amplitude} which, as in non-relativistic quantum mechanics, can be defined in terms of the overlap $\langle f|i\rangle $ between a given initial state $|i\rangle$ and final state $|f\rangle$. The amplitude is a complex number, which makes sense given that the incoming and outgoing particles have a wave-like character, and thus the amplitude must keep track of relative phase differences. If there are different possibilities for the final state, we must add them together to form a total amplitude ${\cal A}$, such that the probability of a given process depends on $|{\cal A}|^2$. This is a real number as it should be, and the fact that different possibilities are added before squaring means that quantum interference effects are correctly included. 

\begin{figure}
    \centering
    \includegraphics[width=0.6\textwidth]{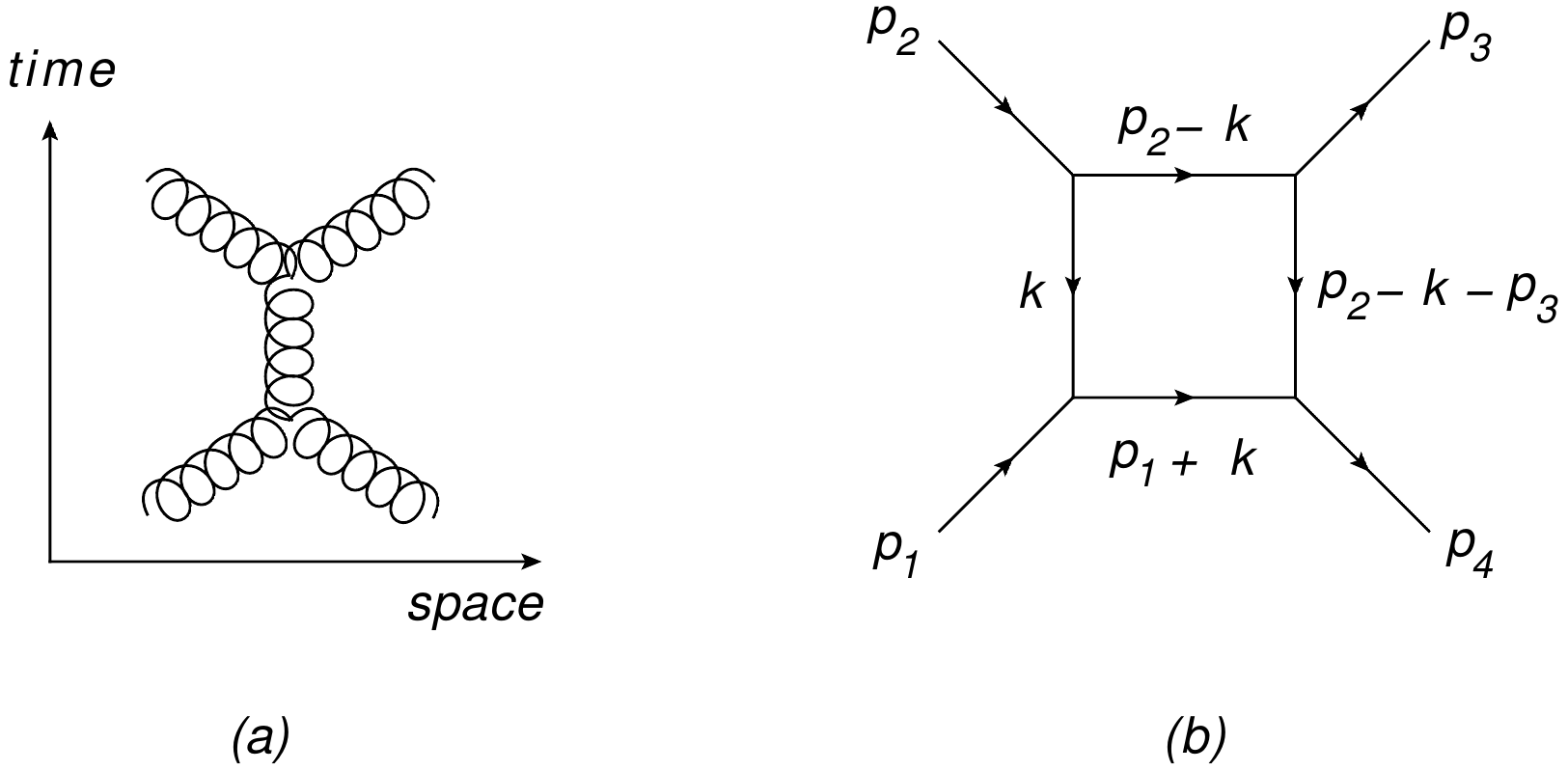}
    \caption{(a) Example Feynman diagram for scattering of two gluons to two gluons; (b) Momentum conservation in a loop diagram.}
    \label{fig:Feynman}
\end{figure}

New methods for calculating amplitudes have blossomed in recent years (see e.g. ref.~\cite{Elvang:2015rqa} for an up-to-date review), but a more traditional way to calculate them is using {\it Feynman diagrams}. These can be viewed as handy space-time pictures that describe how particles can interact, and an example is shown in figure~\ref{fig:Feynman}(a). This shows two gluons (shown as wavy lines) coming together to form a single intermediate gluon, which then decays to two gluons in the final state. More generally, there are different symbols for different types of particle, and they may be coupled together through interaction vertices in prescribed ways.

Feynman diagrams offer much more than a simple way to visualise particle interactions: there are highly precise {\it Feynman rules} that convert each diagram into an algebraic contribution to the scattering amplitude! To go into these rules in detail is clearly beyond the scope of this tutorial, but some of the main ideas are as follows:
\begin{itemize}
\item For a given set of particles in the initial and final states, one must draw all possible diagrams that connect them, with a given number of vertices. 
\item Each vertex is associated with a power of the coupling constant of the the theory ($e$, $g$ or $\kappa$ for electromagnetism, Yang-Mills theory or gravity respectively), plus additional factors involving the momenta of the interacting particles. Note, then, that diagrams with increasing numbers of vertices constitute higher orders in perturbation theory (i.e. an expansion in powers of the coupling).
\item Each line carries a 4-momentum $p^\mu=(E,\vec{p})$, such that 4-momentum is conserved at every vertex. When a diagram involves loops, the external momenta are not sufficient to define all internal momenta. An example is shown in figure~\ref{fig:Feynman}(b), where the lines show the direction of momentum flow. In order to describe the momentum on every line, we have to define the additional {\it loop momentum} $k$, whose possible values must then be summed over according to the rule for combining all possibilities to get the total amplitude. Given that each component of the 4-momentum is a continuous variable, this is a multidimensional integral
\begin{displaymath}
\int\frac{d^4 k}{(2\pi)^4},
\end{displaymath}
where the normalisation is conventional.
\item Internal lines are associated with a factor~\footnote{Note that for intermediate particles, $E$ and $\vec{p}$ are independent degrees of freedom that are not related by the energy-momentum relation of eq.~(\ref{Emom}). Such particles are known as {\it virtual particles} in QFT language, and are not physically measurable directly.}
\begin{equation}
\frac{\ldots}{p^2}\equiv \frac{\ldots}{E^2-\vec{p}^2},
\end{equation}
where the numerator is different for each type of particle.
\item External particles are associated with polarisation vectors or tensors, as appropriate.
\end{itemize}
Calculating amplitudes in a given theory then amounts to knowing what the exact Feynman rules are, and applying them. This is easier said than done: the number of Feynman diagrams increases rapidly with either the number of external particles or loops. Carrying out the integrals over the loop momenta is also highly non-trivial, and involves interesting connections with developing branches of pure mathematics (including special function and number theory). Furthermore, this way of calculating amplitudes makes Yang-Mills theory and gravity look completely different. As an example, the vertex factor for three interacting gluons, written in conventional mathematical notation, has 6 individual terms in it. The 3-graviton vertex instead has over 170 terms! Remarkably, however, a very special structure emerges if we calculate amplitudes in both theories, and compare the results.

Consider a gauge theory amplitude ${\cal A}_m^{(L)}$ for $m$ external particles, and with $L$ loops. The above description of the Feynman rules implies that this will take the general form
\begin{equation}
    {\cal A}_m^{(L)}=
    g^{m-2+2L}\sum_{i} \left(\prod_{l=1}^L\frac{d^4 p_l}{(2\pi)^4}
    \right)\frac{n_i c_i}{\prod_{\alpha_i} p_{\alpha_i}^2}.
    \label{amp}
\end{equation}
Here the sum is over distinct diagrams $i$, where the overall power of the coupling constant is fixed by the number of external particles and loops. There are integrals over the $L$ loop momenta $\{p_l\}$, and each term has a denominator arising from the internal line factors mentioned above. There will be a dependence on the colour charges of the gluons for each diagram (represented by $c_i$). Finally, there is a so-called {\it kinematic numerator} $n_i$ for each diagram, that collects everything else, and which depends on particle momenta and polarisation vectors. Although the general form of eq.~(\ref{amp}) is correct, the kinematic numerators $n_i$ are not unique. For example, gauge transformations (and generalisations of them to include other types of field redefinition) will mix up the numerators associated with different diagrams, such that ${\cal A}_m^{(L)}$ remains invariant. However, it was noticed in 2008~\cite{Bern:2008qj} that there is a particular choice for the numerators $\{n_i\}$ for a given amplitude, that means that they obey similar mathematical identities to the colour factors $c_i$, which arise from the fact that the colour degrees of freedom are described by a Lie group. This itself implies that there must be some mysterious symmetry underlying the kinematic degrees of freedom of a scattering amplitude, and is known as {\it BCJ duality}. It has to this day remained mysterious, although the nature of the symmetry can be glimpsed in certain cases~\cite{Monteiro:2011pc}. However, once the numerators $\{n_i\}$ have been chosen to have this special {\it BCJ-dual form}, it turns out that the formula 
\begin{equation}
    {\cal M}_m^{(L)}=
    \left(\frac{\kappa}{2}\right)^{m-2+2L}\sum_{i} \left(\prod_{l=1}^L\frac{d^4 p_l}{(2\pi)^4}
    \right)\frac{n_i \tilde{n}_i}{\prod_{\alpha_i} p_{\alpha_i}^2}
    \label{amp2}
\end{equation}
describes a gravity amplitude, where $\tilde{n}_i$ is a second set of kinematic numerators, and the gauge theory coupling constant $g$ has been replaced by its gravitational counterpart. Even without knowing what any of the symbols mean, it is clear that eq.~(\ref{amp2}) is almost identical to eq.~(\ref{amp})! The only differences are the replacement of coupling constants, and of colour information by kinematics. This remarkable relationship is the {\it double copy}, and was first presented in refs.~\cite{Bern:2010ue,Bern:2010yg}. 

We have been deliberately imprecise above about which gauge and gravity theories eqs.~(\ref{amp}) and~(\ref{amp2}) correspond to. There are in fact many different types of gauge theory obtained from Yang-Mills by e.g. incorporating additional symmetries such as {\it supersymmetry}, that relates bosonic and fermionic degrees of freedom. One may then take the kinematic numerators $\{n_i\}$ and $\tilde{n}_i$ from different gauge theories, and generate amplitudes in various different gravity theories, which are themselves appropriate generalisations of GR~\footnote{The simplest case of pure Yang-Mills theory copied with itself does not in fact give GR, but GR plus two additional matter particles. We will see the reason for this below.}. It really is amazing that all of these theories should have amplitudes related by the simple replacements above: we stress again that using traditional QFT methods, this structure is entirely invisible.

Above, we started with gauge theory, and replaced colour information by kinematics. We could instead have done the opposite, and obtained the formula
\begin{equation}
    \tilde{\cal A}_m^{(L)}=
    y^{m-2+2L}\sum_{i} \left(\prod_{l=1}^L\frac{d^4 p_l}{(2\pi)^4}
    \right)\frac{c_i \tilde{c}_i}{\prod_{\alpha_i} p_{\alpha_i}^2},
    \label{amp3}
\end{equation}
where we have relabelled the coupling constant to $y$, and introduced a second set of colour factors $\tilde{c}$, that may correspond to a different colour symmetry group in general. This is called the {\it zeroth copy}, and also turns out to yield a scattering amplitude, in so-called {\it biadjoint scalar theory} whose vacuum field equation is:
\begin{equation}
    \partial^2 \Phi^{aa'}-yf^{abc}\tilde{f}^{a'b'c'}\Phi^{bb'}\Phi^{cc'}=0.
    \label{biadjoint}
\end{equation}
In this case, the field $\Phi^{aa'}$ has no spacetime indices (i.e. it is a scalar), but has two different types of colour charge, where the quantities $f^{abc}$ snd $\tilde{f}^{a'b'c'}$ are structure constants in the two different colour symmetry groups. There are no indications that biadjoint scalar theory is a physical theory by itself. However, the above correspondences suggest that at least some of the dynamics of gauge and gravity theories are inherited from the theory of eq.~(\ref{biadjoint}). 

\begin{figure}
    \centering
    \includegraphics[width=0.8\textwidth]{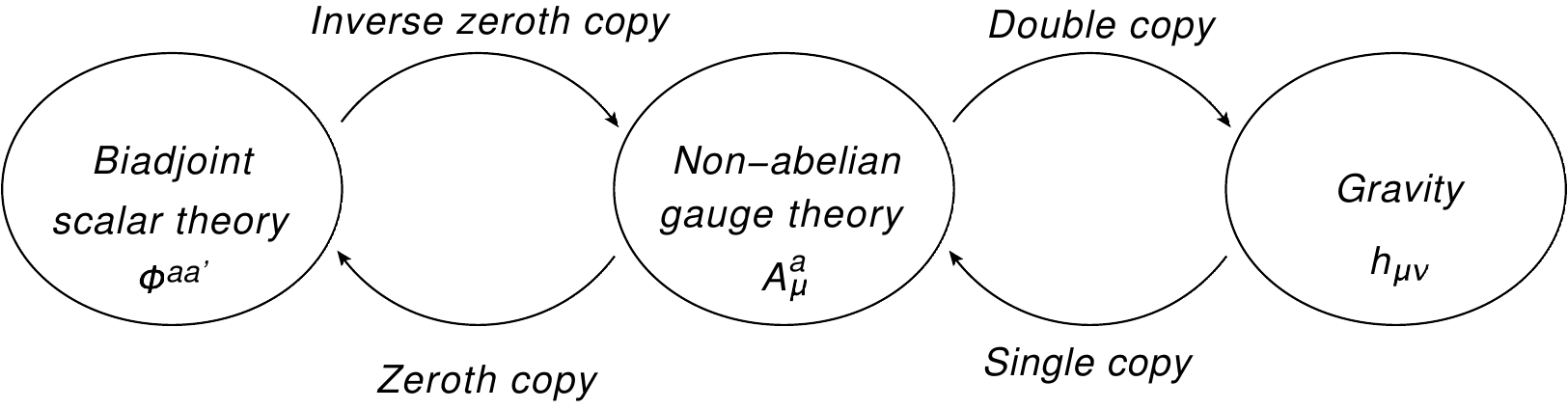}
    \caption{Various theories and the relationships between them. Electromagnetism sits it in the middle, and General Relativity on the right.}
    \label{fig:theories}
\end{figure}

The above discussion has been very technical, so let us now summarise it in more pedestrian terms: scattering amplitudes in a ladder of theories (biadjoint, gauge and gravity) are related by copying correspondences, such that colour charge information in one theory gets replaced by kinematic information (e.g. momenta, polarisations) in another. For the fields themselves, this involves replacing colour charge indices with spacetime indices, and these correspondences are depicted in figure~\ref{fig:theories}. 
The term "double copy" is sometimes used loosely in the literature to refer to any of the types of correspondence in the figure. Furthermore, there is mounting evidence that a whole web of interesting theories is related by double copies, not just the three types of theory considered here~\cite{Bern:2019prr}. We hope it is not too hyperbolic to suggest that the double copy promises to revolutionise our understanding of field theories, given its indication of a deep underlying connection between very different areas of physics. However, for this to be true, we need to know how generally figure~\ref{fig:theories} should be interpreted. Is it merely an accident for scattering amplitudes, or does it apply to the "complete" theories, whatever this means? One way to investigate this is to think of other types of quantity in each theory, and to see whether they can be matched up. A natural starting point is to consider exact classical solutions, and this is the subject of the following section.

\section{The classical double copy}
\label{sec:classical}

In thinking about classical solutions, perhaps the simplest case to think about is that of plane waves, which we have already encountered in gauge and gravity theories in eqs.~(\ref{planewaves}) and~(\ref{hplanewave}). To  double copy the electromagnetic plane wave of eq.~(\ref{planewaves}), we can dress it by an arbitrary constant colour vector $c^a$ to make a Yang-Mills field:
\begin{equation}
    A_\mu^a=c^a A_\mu=c^a \epsilon_\mu e^{ik\cdot x}.
    \label{YMwave}
\end{equation}
Upon substituting this into the Yang-Mills equations, one may show that they linearise~\footnote{Each non-linear term involves a contraction of two colour vectors $c^a c^b$, and structure constants $f^{abc}$. Such contractions vanish due to the antisymmetry of the latter.}, and thus reduce to the Maxwell-like equations
\begin{equation}
    \partial^\mu (\partial_\mu A_\nu^a-\partial_\nu A_\mu^a)
    =c^a \partial^\mu(\partial_\mu A_\nu-\partial_\nu A_\mu)=0.
    \label{MaxwellYM}
\end{equation}
We may thus happily forget about $c^a$ (and hence the Yang-Mills nature of the field), and regard the gauge field as living in electromagnetism only. The photon has two polarisation states, which are often chosen to correspond to the two independent circular polarisations~\footnote{In particle physics language, these are two helicity states of the photon, corresponding to an eigenvalue $\pm 1$ respectively for the projection of the spin along the direction of travel.}
\begin{equation}
    \epsilon^\pm_\mu=\frac{1}{2}(0,\mp 1,i,0).
    \label{pols}
\end{equation}
where the normalisation is such that $\epsilon_+\cdot \epsilon_-=1$. The graviton also has two polarisation states, and one may verify that the appropriate polarisation tensors may be written
\begin{equation}
    \epsilon^\pm_{\mu\nu}=\epsilon^\pm_\mu\epsilon^\pm_{\nu},
    \label{gravpols} 
\end{equation}
such that the gravitational plane waves of eq.~(\ref{hplanewave}) assume the form (for waves of definite helicity)
\begin{equation}
    h^\pm_{\mu\nu}=\epsilon^\pm_\mu \epsilon^\pm_\nu e^{ik\cdot x}.
    \label{hplanewave2}
\end{equation}
Comparing with eq.~(\ref{YMwave}), we see that this has the form of a function that does not change (the phase factor $e^{ik\cdot x}$), times two copies of quantities taken from a gauge theory. Furthermore, one strips away the colour vector in the gauge theory, and replaces it with a kinematic quantity (a polarisation vector) to get the gravity solution. Doing the opposite, one would obtain a biadjoint field 
\begin{equation}
    \Phi^{aa'}=c^a\tilde{c}^{a'}e^{ik\cdot x},
    \label{biadjointwave}
\end{equation}
which is indeed a plane wave solution of eq.~(\ref{biadjoint}). We thus have a series of fields
\begin{displaymath}
    \Phi^{aa'}\quad\leftrightarrow\quad A_\mu^a\quad\leftrightarrow \quad
    h_{\mu\nu},
\end{displaymath}
which provide a realisation of figure~\ref{fig:theories} for particular classical solutions. Indeed, this can be related to the double copy for scattering amplitudes above, in that the incoming and outgoing states in a scattering process are themselves obtained from plane waves. There is also a more explicit similarity with the amplitude story: in going from eq.~(\ref{amp3}) to eqs.~(\ref{amp}) and~(\ref{amp2}), one must successively remove the colour factors and replace them with kinematic numerators, whilst leaving the denominators of the expressions intact. Likewise, in going from eq.~(\ref{biadjointwave}) to eqs.~(\ref{YMwave}) and~(\ref{hplanewave2}), one must strip off the colour vectors and replace them with polarisation vectors, leaving a certain function ($e^{ik \cdot x}$) alone. The latter therefore seems to play an analogous role to the denominator factors in an amplitude, and indeed this analogy can be made more  precise~\cite{Monteiro:2014cda}. 

Note that particular combinations of polarisation states have been taken in eq.~(\ref{hplanewave2}), but that there are two other choices: in combining the two polarisation states of two photons, we expect four different combinations. For the missing two, we can take them to be the (anti-)symmetric combinations
\begin{equation}
    \frac12\left[\epsilon^+_\mu\epsilon^-_\nu-\epsilon^-_\mu\epsilon^+_\nu
    \right],\quad \frac12\left[\epsilon^+_\mu\epsilon^-_\nu+\epsilon^-_\mu\epsilon^+_\nu
    \right].
    \label{polstates}
\end{equation}
Each of these constitutes a single degree of freedom, and the first and second combinations can be associated with a (pseudo)-scalar field respectively~\footnote{A pseudo-scalar field differs from a scalar in that it picks up a minus sign under a {\it parity} transformation, that reverses the spatial coordinate axes.}. These are known as the {\it axion} and the {\it dilaton}, and thus the true double copy of pure Yang-Mills theory is not General Relativity, but gravity coupled to these two extra fields. There are additional justifications one can give for this result. For 
Feynman diagrams with no loops, it turns out that the double copy can be derived from string theory, where it reproduces known results~\cite{Kawai:1985xq}, and also makes clear that the axion and dilaton should be present. This argument does not, however, easily generalise to amplitudes with loops. However, taking a "product" of gauge fields, on general grounds, generates a field with two indices, which can be decomposed into its symmetric traceless, antisymmetric, and trace degrees of freedom. These are known to mathematicians as {\it irreducible representations of the Lorentz group}, and correspond physically to the graviton, axion and dilaton respectively. 

The question naturally arises as to whether the above procedure can be extended, and indeed it can. The double copy of more general exact classical solutions was first considered in ref.~\cite{Monteiro:2014cda}, which considered the infinite family of so-called {\it Kerr-Schild} solutions in GR. For each case, the metric can be written as in eq.~(\ref{hdef}), with the graviton having the special form
\begin{equation}
    h_{\mu\nu}=\phi k_\mu k_\nu,
    \label{hKS}
\end{equation}
where $\phi$ is a scalar field, and the vector field $k_\mu$ has to satisfy
\begin{equation}
    k^2=0,\quad k\cdot \partial k^\mu=0.
    \label{KSconditions}
\end{equation}
Substituting this ansatz into the Einstein equations of GR, they are found (after a great deal of effort) to linearise. Thus, they become much easier to solve, and furthermore any solutions are then known to be exact. Looking at eq.~(\ref{hKS}), it is tempting to guess how to take its single copy and zeroth copies. One could simply remove successive factors of the 4-vector $k_\mu$, and replace them with colour vectors to get
\begin{equation}
    A_\mu^a=c^a \phi k_\mu,\quad \Phi^{aa'}=c^a\tilde{c}^{a'}\phi.
    \label{singlezeroth}
\end{equation}
Reference~\cite{Monteiro:2014cda} proved that, for static solutions at least, the fields of eq.~(\ref{singlezeroth}) are solutions of the Yang-Mills and biadjoint equations respectively, where the field equations are also linearised in each case. Some time-dependent cases are also known, including the plane waves already mentioned above! Interestingly, the Kerr-Schild double copy involves solutions of pure GR, without any contamination from the axion and dilaton. To see this, note that the field of eq.~(\ref{hKS}) is symmetric in its indices by construction, and also traceless due to the conditions on $k^\mu$ of eq.~(\ref{KSconditions}). It thus corresponds to the graviton only, and no product of gauge fields involving $k_\mu$ would be able to generate the axion (antisymmetric) or dilaton (trace degree of freedom). 

The Kerr-Schild double copy may look abstract, but it in fact includes some of the most famous gravitational objects known to physics! A first example is the Schwarzschild black hole, a static, spherically-symmetric solution which can be sourced by a pointlike mass $M$ at the origin. Its Kerr-Schild scalar field and vector are found to be
\begin{equation}
    \phi=\frac{M}{4\pi r},\quad k^\mu=(1,\vec{e}_r),
    \label{Schwarzschild}
\end{equation}
where $r$ is the spherical radius, and $\vec{e}_r$ a unit 3-vector in the radial direction. Taking the single copy results in the gauge field
\begin{equation}
    A_\mu=\frac{Q}{4\pi r}k^\mu\quad\rightarrow\quad
    \left(\frac{Q}{4\pi r},0,0,0\right),
    \label{pointcharge}
\end{equation}
where we have performed a gauge transformation on the right-hand side, whose details may be found in ref.~\cite{Monteiro:2014cda}. Recalling that the zeroth component of the gauge field is the electrostatic potential, we recognise eq.~(\ref{pointcharge}) as the gauge field of a point charge. Thus, the double copy relates a simple point charge at the origin to a point mass in gravity, which is directly analogous to the replacement of colour information by kinematics in the double copy for amplitudes! 

\begin{figure}
    \centering
    \includegraphics[width=0.3\textwidth]{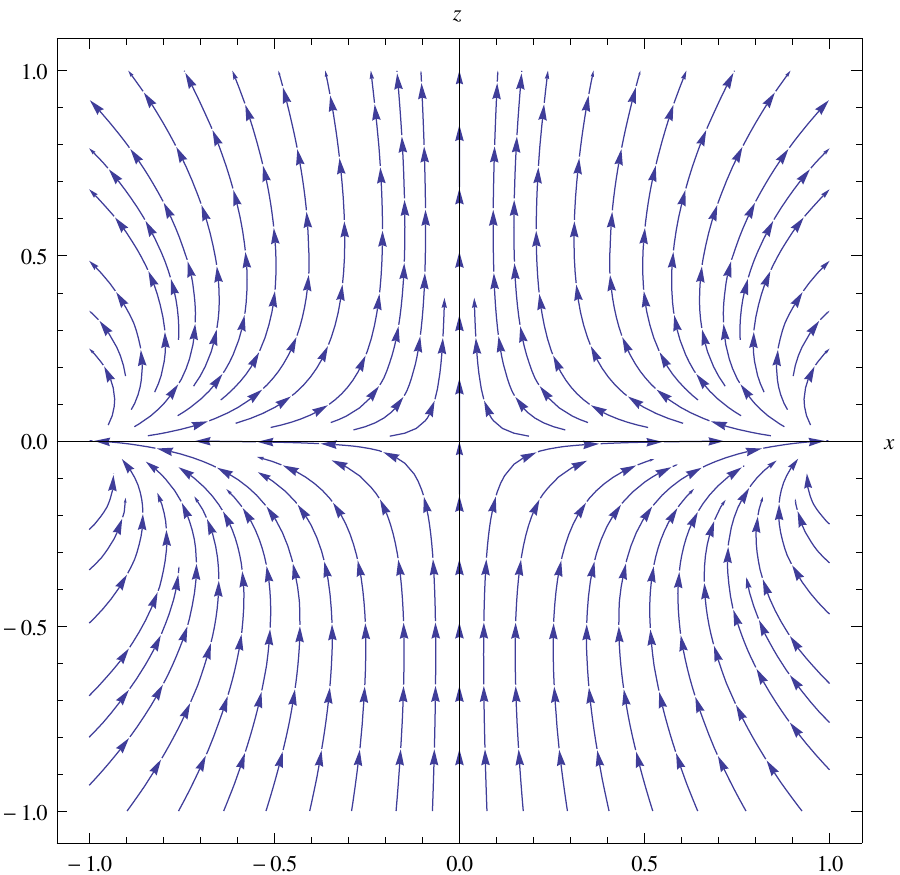}
    \includegraphics[width=0.3\textwidth]{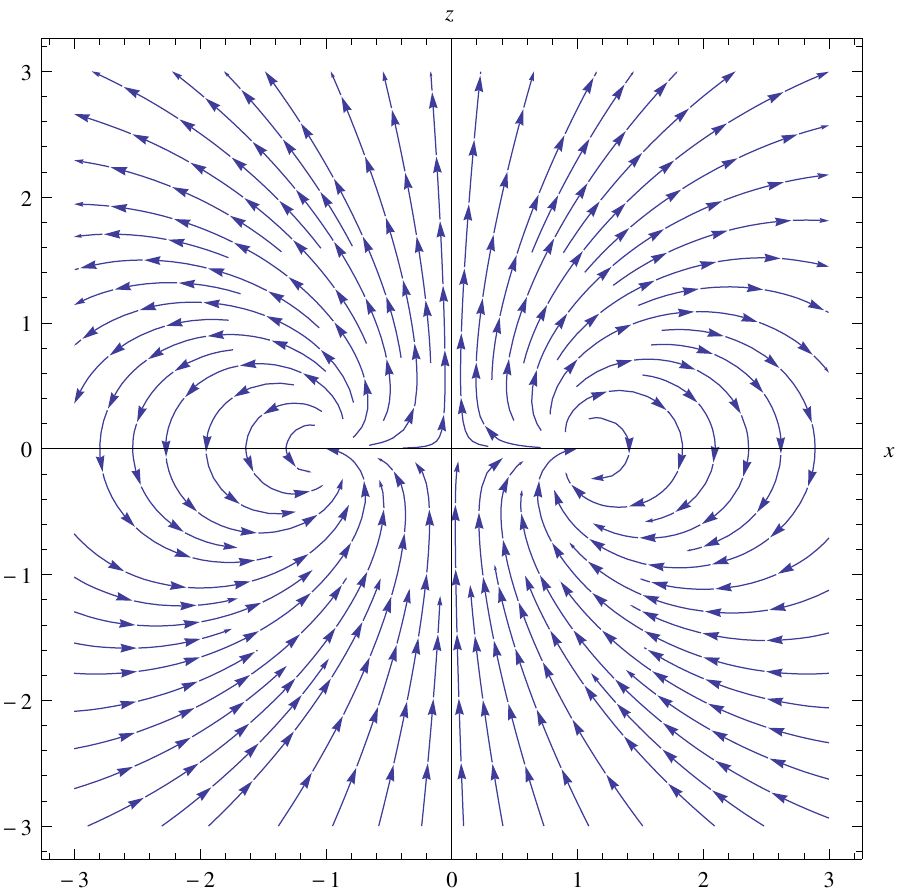}
    \includegraphics[width=0.3\textwidth]{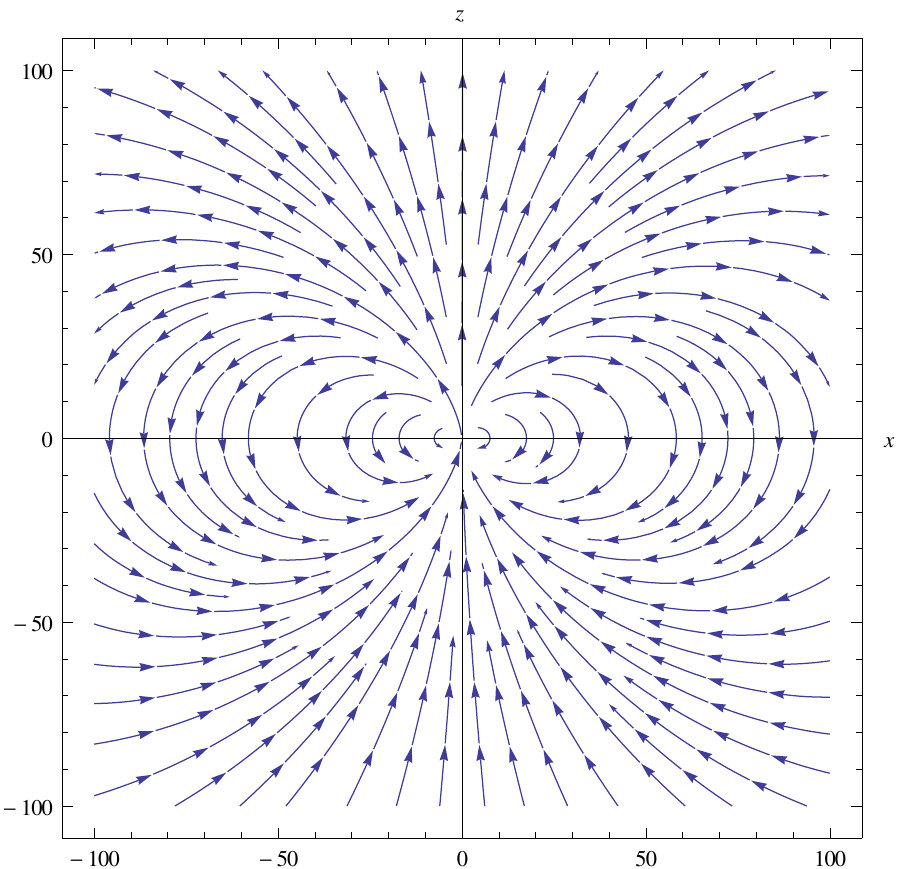}
    \caption{The magnetic field of the single copy of the Kerr (rotating) black hole, corresponding to a rotating disc of charge. The disc is seen sideways on, and runs from $x=-1$ to $x=1$.}
    \label{fig:magfield}
\end{figure}

As a more non-trivial example, one may consider the {\it Kerr black hole}, an axially symmetric system which can be sourced by a rotating disc of mass in gravity. Similar to above, its single copy turns out to be a rotating disc of charge, whose profile matches that of its gravitational counterpart. In figure~\ref{fig:magfield}, we show the magnetic field of the single copy of the Kerr black hole. As we zoom out, the field becomes dipole-like, as one expects given that a rotating disc of charge looks like a nested set of current loops!

Other examples of the Kerr-Schild double copy involve magnetic monopoles~\cite{Luna:2015paa}, accelerating particles~\cite{Luna:2016due}, electromagnetic vortices~\cite{Ilderton:2018lsf} and novel solutions in two spatial dimensions~\cite{CarrilloGonzalez:2019gof}. However, the family of solutions that can be put into a Kerr-Schild form remains very special, so that it would be desirable to extend the scope of the classical double copy yet further. To this end, an alternative procedure known as the {\it Weyl double copy} has been defined~\cite{Luna:2018dpt}, that relies on the so-called {\it spinorial formalism} of field theory (reviewed both excellently and encyclopaedically in refs.~\cite{Penrose:1987uia,Penrose:1986ca}). This can indeed be made more general than the Kerr-Schild double copy~\cite{Godazgar:2020zbv,White:2020sfn,Chacon:2021wbr}, and has also been explored in the conventional tensor language~\cite{Alawadhi:2020jrv}. Another deficiency of the Kerr-Schild approach is that it is highly dependent on a particular gauge (coordinate system) being chosen in gauge theory (gravity), which itself is reminiscent of the fact that the double copy for amplitudes only works for a particular {\it generalised gauge} such that the kinematic numerators $\{n_i\}$ are BCJ-dual. Alternative formalisms have been developed that can in principle work in any gauge~\cite{Anastasiou:2014qba,Anastasiou:2018rdx,Borsten:2020xbt,Borsten:2020zgj,Borsten:2021hua,Campiglia:2021srh}, where the "unphysical" degrees of freedom in each theory (i.e. corresponding to gauge redundancy in the fields) can themselves be matched up.

It is not yet known whether a fully general and exact statement of the double copy can be made, that works for {\it any} type of solution, be it classical or quantum. Attempts to generalise the idea include finding (and trying to copy)  non-linear solutions of the biadjoint theory~\cite{White:2016jzc,DeSmet:2017rve,Bahjat-Abbas:2018vgo,Bahjat-Abbas:2020cyb}, showing how exact symmetries in different theories can be related~\cite{Alawadhi:2019urr,Banerjee:2019saj,Huang:2019cja}, and looking at how {\it topological} properties of solutions can be matched up~\cite{Berman:2018hwd,Alfonsi:2020lub}. If one is willing to forego exact statements and work order-by-order in perturbation theory, there are by now many different approaches for calculating classical observables in General Relativity by first calcuating in a (simpler) gluon theory, and then double copying the results (see e.g. ref.~\cite{Bern:2019prr} and references therein). This is continuing to attract intense worldwide interest due to the applications to gravitational waves. A typical signal observed by the LIGO experiment arises from the coming together of two heavy objects such as black holes or neutron stars, in a three stage process: (i) an {\it inspiral phase}, in which the objects gradually orbit closer to each other; (ii) the {\it merger} itself, in which e.g. two black holes combine to make a larger one; (iii) the {\it ringdown} phase, in which the combined heavy object wobbles and settles down (see ref.~\cite{Levi:2018nxp} for an excellent review of contemporary methods for describing such processes). The double copy has so far been applied to step (i), and may potentially be used for step (iii). Step (ii) relies on complex numerical simulation work, but the expense of this may be reduced if further improvements can be made to the other two steps.

\section{What's in it for optics?}
\label{sec:optics}

One of the key scientific problems of our age is a tendency towards increasing specialisation and isolation. Physicists start to significantly diverge from each other even before they have completed their Masters degrees, and the lack of a common language between apparently diverse fields such as astrophysics / cosmology, high energy physics, condensed matter and optics too often impedes the ability of different types of physicist to work together. In this article, we have reviewed an intriguing set of correspondences -- known collectively as the {\it double copy} -- that relate solutions in widely different field theories. Unlike many of the developments that have preoccupied theoretical high energy physicists in recent years, the double copy connects actual physical theories that are directly observed in nature! It is thus only natural to try to widen the notion of the double copy yet further, and to seek situations that may be tested in the lab. It is difficult for the present author to definitively state what these experiments might be, especially given that the aim of this article is to try to stimulate the kinds of conversations that might lead to the necessary interdisciplinary work. However, there are some potentially useful connections to optics / condensed matter that have appeared in recent years.

Reference~\cite{Fernandez-Corbaton:2014cha} showed that classical gravitational waves could in principle be emulated using quantum-entangled photon pairs. It was shown how to obtain the appropriate gravitational wave polarisations by combining definite photon helicity states, but it was also noted that two of the possible combinations were associated with helicity zero, and thus had to be thrown away. This is in fact the double copy in all but name, and the two "spurious" photon combinations are precisely the axion and dilaton we discussed above! Rather than being discarded, they could be used to simulate the full double copy of a pure gauge theory. Regarding what to simulate, ref.~\cite{Fernandez-Corbaton:2014cha} had in mind the emulation of gravitational waves moving through a background curved spacetime, where the latter could be modelled by passing the entangled photons through a metamaterial. Unknown to the authors was the fact that whether or not the double copy can be be made exact in a curved spacetime is in fact an open problem (see e.g. refs.~\cite{Bahjat-Abbas:2017htu,Carrillo-Gonzalez:2017iyj,Prabhu:2020avf,Alkac:2021bav,Borsten:2021zir} for preliminary investigations), that in turn may shed light on how the double copy may be applied in astrophysics and / or cosmology. Further studies have involved scattering particles on localised wave backgrounds, which could be simulated using e.g. strong laser pulses~\cite{Adamo:2017nia}. It is highly exciting that the possibility exists of using table-top experiments to probe one of the most intriguing ideas to emerge in high energy physics in recent years!

Another connection noticed by the present author is that of the study of topologically non-trivial states of light. A family of solutions of the Maxwell equations known as {\it Hopfions} have electric or magnetic field lines that are knotted, and generalisations exist such that the field lines form so-called {\it torus knots}. Gravitational counterparts also exist, and have been studied in refs.~\cite{Dalhuisen:2012zz,Swearngin:2013sks,deKlerk:2017qvq,Thompson:2014owa,Thompson:2014pta,Sabharwal:2019ngs}, where the latter reference comments that the EM and gravity solutions can be related by the Weyl double copy mentioned above. The authors use {\it twistor theory}~\cite{Penrose:1967wn,Penrose:1972ia,Penrose:1968me} -- an elegant set of mathematical ideas relating algebraic geometry and complex analysis -- to classify knotted radiation solutions in a compact way. This was taken further in refs.~\cite{White:2020sfn,Chacon:2021wbr}, which provided a derivation of the Weyl double copy using twistor ideas (see also refs.~\cite{Elor:2020nqe,Farnsworth:2021wvs} for related ideas). It would be interesting to know whether experimental efforts exist to study the electromagnetic knotted solutions in their own right. If so, might they be combined with the above ideas to simulate gravitational knotted radiation? What could this teach us?

\section{Conclusion}
\label{sec:conclusion}

Field theories occur in many branches of physics, and notable examples include the gauge theories underlying three of the four fundamental forces of nature (including electromagnetism), and General Relativity. Recent years have seen a flurry of activity regarding a newly discovered correspondence between gauge and gravity theories, called the {\it double copy}. Originating in the study of scattering amplitudes related to the probability for particles to interact, the double copy has since been extended to classical solutions, including those of interest in astrophysics and optics. It has also been shown to apply to an increasing range of theories, with varying degrees of physical relevance. 

As the remit of the double copy has increased, so have the number of interested researchers from astrophysics, cosmology, high energy physics and pure mathematics. Now seems to be the perfect time to try to interest physicists from optics and / or condensed matter, particularly given recent indications that the scope and applicability of the double copy may be amenable to table-top experiments in the lab.

The author humbly hopes that this article will stimulate interested conversations -- and indeed conversations about how to have the conversations -- that are able to surmount the often considerable barriers between different subfields, which are often merely due to a difference in language. Recent years have taught us that our traditional approach to quantum field theory is hiding a deep and profound underlying structure. Will the field of optics shine the light that guides the way?

\section*{Acknowledgments}

I thank Johannes Courtial and Kurt Busch for their encouragement in
writing this article. This work has been supported by the UK Science
and Technology Facilities Council (STFC) Consolidated Grant
ST/P000754/1 ``String theory, gauge theory and duality'', and by the
European Union Horizon 2020 research and innovation programme under
the Marie Sk\l{}odowska-Curie grant agreement No. 764850 ``SAGEX''.


\bibliography{refs}

\end{document}